\documentclass[preprint,aps,prc,showpacs,nofootinbib]{revtex4}
\usepackage{epsfig}
\usepackage{graphicx,amsmath}

\newcommand\ba{\begin{eqnarray}}
\newcommand\ea{\end{eqnarray}}
\newcommand\nn{\nonumber}
\newcommand{\be}{\begin{equation}}
\newcommand{\ee}{\end{equation}}

\begin{document}
\title{Study of resonant processes for multi-pion production in 
$\bar p +p\to\ell ^++\ell^-  +n_\pi \pi$ annihilation }

\author{E. A. Kuraev}

\altaffiliation{\it JINR-BLTP, 141980 Dubna, Moscow region, Russian
Federation}
\affiliation{ DAPNIA/SPhN, CEA/Saclay, 91191 Gif-sur-Yvette Cedex, France}

\author{C. Adamu\v s\v c\'in }
\altaffiliation{\it Department of Theoretical Physics, IOP, Slovak Academy of Sciences, Bratislava, Slovakia}
\affiliation{ DAPNIA/SPhN, CEA/Saclay, 91191 Gif-sur-Yvette Cedex, France}

\author{E. Tomasi-Gustafsson}
\email{etomasi@cea.fr}
\affiliation{ DAPNIA/SPhN, CEA/Saclay, 91191 Gif-sur-Yvette Cedex, France}

\author{F. Maas}
\affiliation{CNRS/IN2P3, Institut de Physique Nucl\'eaire, UMR 8608 \\and
Univ. Paris-Sud, Orsay, F-91405 France}

\date{\today}

\begin{abstract}
In frame of a phenomenological approach based on Compton-like Feynman amplitudes, we study multi-pion production in antiproton nucleon collisions. The main interest of this reaction is related to the possibility to study the properties of the presumable $\bar N N$ atom  and of its resonances. For the case of formation of a scalar or pseudoscalar resonant state, with
$I^G(J^{P})=1^-(0^{\pm}),~0^+(0^-)$  numerical results are obtained. The differential cross section in an experimental set-up where the pions invariant mass is measured, is given with explicit dependence on the lepton pair and pions invariant mass. 
\end{abstract}

\maketitle

In a previous paper \cite{Ad06} we calculated the differential cross section for single pion production in the reactions $\bar p+N\to \pi +\ell^+ + \ell^-$, and showed that such reactions are measurable at upcoming facilities, bringing unique information on electromagnetic and axial nucleon form factors in the unphysical region of  time-like momentum transfer squared, $q^2$.

The purpose of the present work is to study the resonant production of a number $n_{\pi}$ of  pions accompanied by a lepton pair, in the antiproton nucleon annihilation process. We limit our discussion to quantum numbers of the excited resonances: $I^G(J^P)=1^+(0^\pm )$ and $0^+(0^\pm)$. The existence a $\bar N N $ bound state and of its resonances was predicted in a series of papers, see Ref. \cite{Bo74,An01}. A partial wave analysis performed in \cite{An01} in the range $1960<M^*<2410$ combines data in different decay channels $3\pi^0$, $\pi^0\eta $, $\pi^0\eta \eta' $ and reveals a series of resonant states in the $p\bar p$ system. Some of the states were precisely identified, in particular two states with quantum numbers $I(J^{PC})=1(0^{-+ })$, with masses $M^*=2360$ and 2070 MeV, respectively and widths about 300 MeV. Higher spin states were also identified. Such states show similar masses as dibaryon states, except that they exhibit large widths, which are explained by their decay through hadronic states, which is impossible for dibaryons \cite{Tati}.

$N\bar N$ state has hydrogen-like atomic structure. Similarly to positronium, in addition to strong interaction, QED effects are present, due to the large masses (small distances) involved. Special attention is payed to resonances with masses $\sim 2M$, where $M$ is the nucleon mass.  The resonances with zero orbital momentum are expected to have large width, whereas higher values of orbital momentum lead to a smaller width given by \cite{Bo74}:
\be
 \Gamma\simeq \frac {100}{[(2\ell +1)!!]^2} \mbox{~MeV~}, 
\label{eq:eq1}
\ee
which gives, for example, $\Gamma=100$ MeV for $\ell=0$,  $\Gamma=10$ MeV for $\ell=1$, and   $\Gamma=0.5$ MeV for $\ell=2$. 

Experimental evidence of such series of dibaryon resonances exist \cite{Tati}, although it is still controversial. An investigation of this problem will be possible at the upcoming FAIR facility \cite{FAIR}, where the planned antiproton beams render possible the  measurement of the reactions investigated in the present paper.

The emission of a lepton pair permits to select the appropriate kinematics 
adapted to the excitation of  such  resonances, when the total $\bar pN$ center of mass (CMS) energy exceeds the mass of the resonances, due to a known mechanism, called 'return to resonances' \cite{Ba68}: the lepton pair carries away the extra energy and momentum, providing the condition of exciting the resonances. Experimentally it manifests through a deformation of the Breit Wigner distribution, yielding a 'radiative' tail. For the excitation of a narrow resonance, the emission of soft quanta alters the shape of the resonance curve: the height of the resonance curve decreases and a radiative tail arises. Indeed, for total energy $W$, higher than the resonance mass $M'$, the most favourable mechanism is when the energy "excess" $W-M'$ is absorbed by the photon emission (with energy
$\omega=W-M'$). Here the small extra factor $\alpha$  is compensated by the resonance denominator.

\begin{figure}
\begin{center}
\includegraphics[width=12cm]{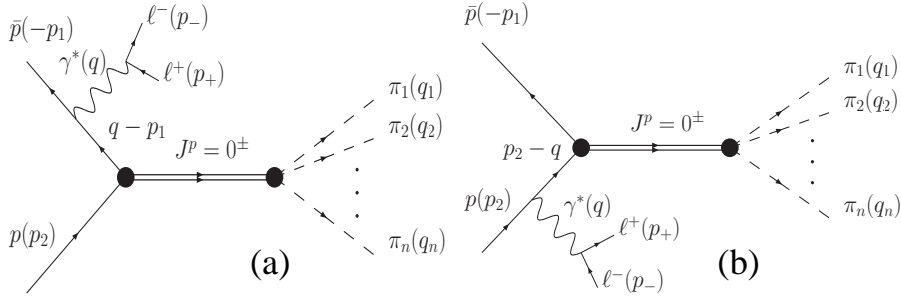}
\caption{\label{fig:fig1} Feynman diagrams for the reaction $ \bar{p} +p \to \ell^++\ell^-+n_{\pi}\pi$: lepton pair emission form the antiproton (a) and from the proton  (b).}
\end{center}
\end{figure}

Let us consider the annihilation of an antiproton and a proton in a lepton pair, and a resonant state which subsequently decays into a set of $n_{\pi}$ pions: $\bar p(p_1) +p(p_2) \to \gamma^*(q) + R(J^P,M')\to  \ell^+ (p_+) +\ell^-(p_-) + \pi_1(q_1)+...+\pi_{n_{\pi}}(q_{n_{\pi}})$. Such state can have the quantum number of scalar, pseudoscalar, axial or vector particle. The Feynman diagrams for such reactions are plotted in 
Fig. \ref{fig:fig1}. We neglect direct photon emission from the resonant state, assuming that it is (indirectly) included $via$ the width of the resonance.

To describe the kinematics of the process it is convenient to introduce two kinematical variables: the invariant mass squared of the lepton pair 
\be
q^2=(p_+ +p_-)^2
\label{eq:eqq}
\ee
and the one of the pions:
\be
s_1=(\sum_i q_i)^2=(p_1+p_2-q)^2.
\label{eq:eqs}
\ee
One can write the phase volume as:
\ba
\int d\Gamma_x&=&\int (2\pi)^{-2}\delta^4\left (p_1+p_2-p_+-p_- -\sum_{i=1}^x q_i\right )
\frac{d^3p_+}{2\epsilon_+}\frac{d^3p_-}{2\epsilon_-}\Pi_{i=1}^x
\frac{d^3q_i}{2\epsilon_i(2\pi)^3}\nonumber\\
&=&\int d\Gamma_q\int d\Gamma_Q^{(n_{\pi})}\int d\Gamma_ {Qq}
\ea
with
$$
\int d\Gamma_q=\int \frac{d^3p_+}{2\epsilon_+ (2\pi)^3}\frac{d^3p_-}{2\epsilon_-(2\pi)^3}\delta^4 (q-p_+ -p_-),~
$$
$$
\int d\Gamma_Q^{(n_{\pi})}=\int\Pi_{i=1}^{n_{\pi}}
\frac{d^3q_i}{2\epsilon_i(2\pi)^3}\delta^4\left (Q-\sum_{i=1}^{n_{\pi}} q_i\right )(2\pi)^4,
$$
$$
\int d\Gamma_ {Qq}=\int d^4q d^4Q \delta^4(p_1+p_2-q-Q).
$$
It is convenient to calculate this last quantity in CMS, where $\vec q=-\vec Q$, $\vec p_1=- \vec p_2$. In this reference frame we have:
\be
\int d^4q=\frac{1}{2}dq_O |\vec q| dq^2 dO_q =\int d\Gamma_{Qq},
\ee
where $dO_q =2\pi d\cos\theta_q$ is the element of solid angle, integrated on the azymuthal angel $\phi$. Then, using the relations $\sqrt{s}=q_0+Q_0;~(s_1-q^2)/\sqrt{s}=Q_0-q_0$,  and $s=(p_1+p_2)^2$,  one finds: 
\be
\int d\Gamma_{Qq}=\frac{\pi}{2}\int_{n_{\pi}m_{\pi}}^{\sqrt{s}-\sqrt{q^2}} \frac{ds_1}{s}dq^2d\cos \theta_q
\sqrt{\Lambda(s,s_1,q^2)},~\Lambda(a,b,c)=a^2+b^2+c^2-2(ab+ac+bc), 
\ee
with $\theta_q=\widehat{\vec q,\vec p_1}$. The function $\Lambda(s,s_1,q^2)$ must be positive in the physical region.

Let us restrict our considerations to scalar $J^P=0^+$  and pseudoscalar $J^P=0^-$ states.  The corresponding matrix elements are:
\be
{\cal M}^i_{res}=\frac {4\pi\alpha}{q^2}V_{\rho}(q^2){\cal J}_{\mu}
\bar v(p_1)
{\cal O}_{\mu}^i
u(p_2) \displaystyle\frac{g_1^ig_2}{s_1 -M'^2+iM'\Gamma '}
\label{eq:eqm}
\ee
where  $M'$ and $\Gamma '$ are the mass and the width of the resonance, 
$V_{\rho}(q^2)=m_\rho ^2/(m_\rho^2-q^2)$ corresponds to the coupling of the photon to the $\rho$ meson ($m_\rho $ is the $\rho$ meson mass), when interacting with a nucleon, ${\cal  J}_\mu(q)=\bar{v}(p_+)\gamma_\mu u(p_-)$ is the leptonic current
and ${\cal O}_{\mu}^i$ 
\be
{\cal O}_{\mu}^i=
 {\cal F} _{\mu}^p(q)
\displaystyle\frac{-\hat p_1+\hat q +M}{(p_1-q)^2-M^2}\gamma^{(i)} 
 + \displaystyle\frac{\hat p_2-\hat q +M}{(p_2-q)^2-M^2}
{\cal F} _{\mu}^p(q)\label{eq:eqo}
\ee
(where $\gamma^{(i)}=\gamma_5$  for a pseudoscalar resonant state and $\gamma^{(i)}=1$ for a scalar state) contains the hadronic structure:
\be
{\cal F}_{\mu}^p(q) =
F_1^p(q^2)\gamma_\mu+\frac{F_2^p(q^2)}{4M}(\hat{q}\gamma_\mu-\gamma_\mu\hat{q}),
\label{eq:eqff}
\ee
which is parametrized in terms of the Pauli and Dirac proton form factors, $F_1$ and $F_2$ (in time-like region).

Let us note that the hadron current 
${\cal  J}_\mu^h=\bar v(p_1){\cal O}_{\mu}^iu(p_2)$ obeys the gauge invariance: ${\cal  J}_\mu^h q_\mu=0$.

In Eq. (\ref{eq:eqm}), the coupling constant $g_2$ is related to the width of the decay channel:
\be
\Gamma '_{n_{\pi}}=\frac{1}{2M'} g_2^2 \int d \Gamma_Q^{({n_{\pi}})}, 
\ee
where $\Gamma '_{n_{\pi}}$ is the width of the decay of the $p\bar p$ state to $n_{\pi}$ pions, $g_1$ represents the coupling constant at the $\bar p p$ vertex to the resonant state. When the resonance is heavier than twice the  nucleon mass its value can be related to the $\bar p p$ decay width:
\be
(g_1^i)^2=   \displaystyle\frac{ 8 \pi\Gamma_{\bar p p}  }{M'A_i}, 
\ee
with $ A_i =\beta$ for $J^P=0^-$, and $ A_i =\beta^3 $ for $J^P=0^+$ and 
$\beta=\sqrt{1-(4M^2/M'^2)}$.

\begin{figure}
\begin{center}
\includegraphics[width=8cm]{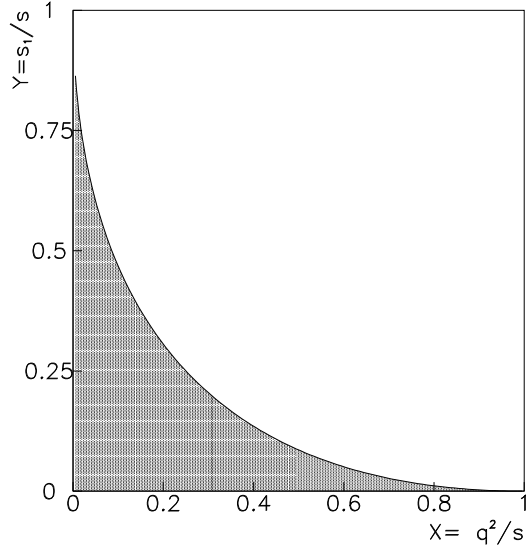}
\caption{\label{Fig:kin} Allowed kinematical region, as a function of $X=q^2/s$ and $Y=s_1/s$ (shaded area).}
\end{center}
\end{figure}

The corresponding differential cross sections are:
\be
\displaystyle\frac{d \sigma^i}{dq^2 ds_1}=
\displaystyle\frac{2\alpha ^2}{3\pi}
\displaystyle\frac{\Gamma '_{n_\pi} \Gamma ^i_{\bar p p}}
{(s_1-M'^2)^2+M'^2 \Gamma_i^2}
\displaystyle\frac{\sqrt{\Lambda (s,s_1,q^2)}}{s^2r}
\displaystyle\frac{1}{q^2} {\cal D}_i,~{for~}M'>2M,
\label{eq:eqss}
\ee
and 
\be
\displaystyle\frac{d \sigma^i}{dq^2 ds_1}=
\displaystyle\frac{\alpha ^2M'\Gamma '_{n_\pi}}{12\pi^2 q^2}
\displaystyle\frac{ (g_i)^2}
{(s_1-M'^2)^2+M'^2 \Gamma_i^2}
\displaystyle\frac{\sqrt{\Lambda (s,s_1,q^2)}}{s^2r}
{\cal D}_i,~{for~}M'<2M,
\label{eq:eqss1}
\ee
with $r=\sqrt{1-(4M^2/s)}$ and
\be
{\cal D}^i=\int  d \cos \theta_q 
\left (g_{\mu\nu} -
\frac{q_{\mu}q_{\nu}}{q^2} \right ) S^{(i)}_{\mu\nu} ,
\label{eq:eqdi}
\ee
and
\be
S^{(i)}_{\mu\nu} = \displaystyle\frac{1}{4} Tr (\hat p_1-M){\cal O}_{\mu}^i (\hat p_2+M) {\cal O}_{\nu}^{i*}.
\label{eq:eqds}
\ee
Two dimensionless variables,  $X=q^2/s$ and $Y=s_1/s$ can been defined, using the invariant mass of the lepton pair, $q^2$, Eq. (\ref{eq:eqq}), and the invariant mass of the pions, $s_1$, Eq. (\ref{eq:eqs}). The kinematical region scanned by the reaction, is defined by the conditions $\Lambda(s,s_1,q^2)>0$ and it corresponds to $Y\le (1-\sqrt{X})^2$, as it is shown in Fig. \ref{Fig:kin}.

The angular integration entering in $D^i$, Eq. (\ref{eq:eqdi}), can be calculated with the help of the following relations:
\ba
\int \displaystyle\frac{d\cos\theta_q }{d_1^2}&=&
\int \displaystyle\frac{d\cos\theta_q }{d_2^2}
=\displaystyle\frac{2s}{q^2ss_1+M^2\Lambda(s,s_1,q^2)};
\nn \\
\int \displaystyle\frac{d\cos\theta_q }{d_1}&=&
\int \displaystyle\frac{d\cos\theta_q }{d_2}
= \displaystyle\frac{2}{r\sqrt{\Lambda(s,s_1,q^2)}}
\log \left |
\displaystyle\frac{q^2+s_1-s+r\sqrt{\Lambda(s,s_1,q^2)}}
{q^2+s_1-s-r\sqrt{\Lambda(s,s_1,q^2)}}\right |,
\ea
with  $d_{1,2}=q^2-2p_{1,2}q$ and $d_1+d_2=s_1+q^2-s$. The integration is performed  in the center of mass of the initial particles. 

\begin{figure}
\begin{center}
\includegraphics[width=12cm]{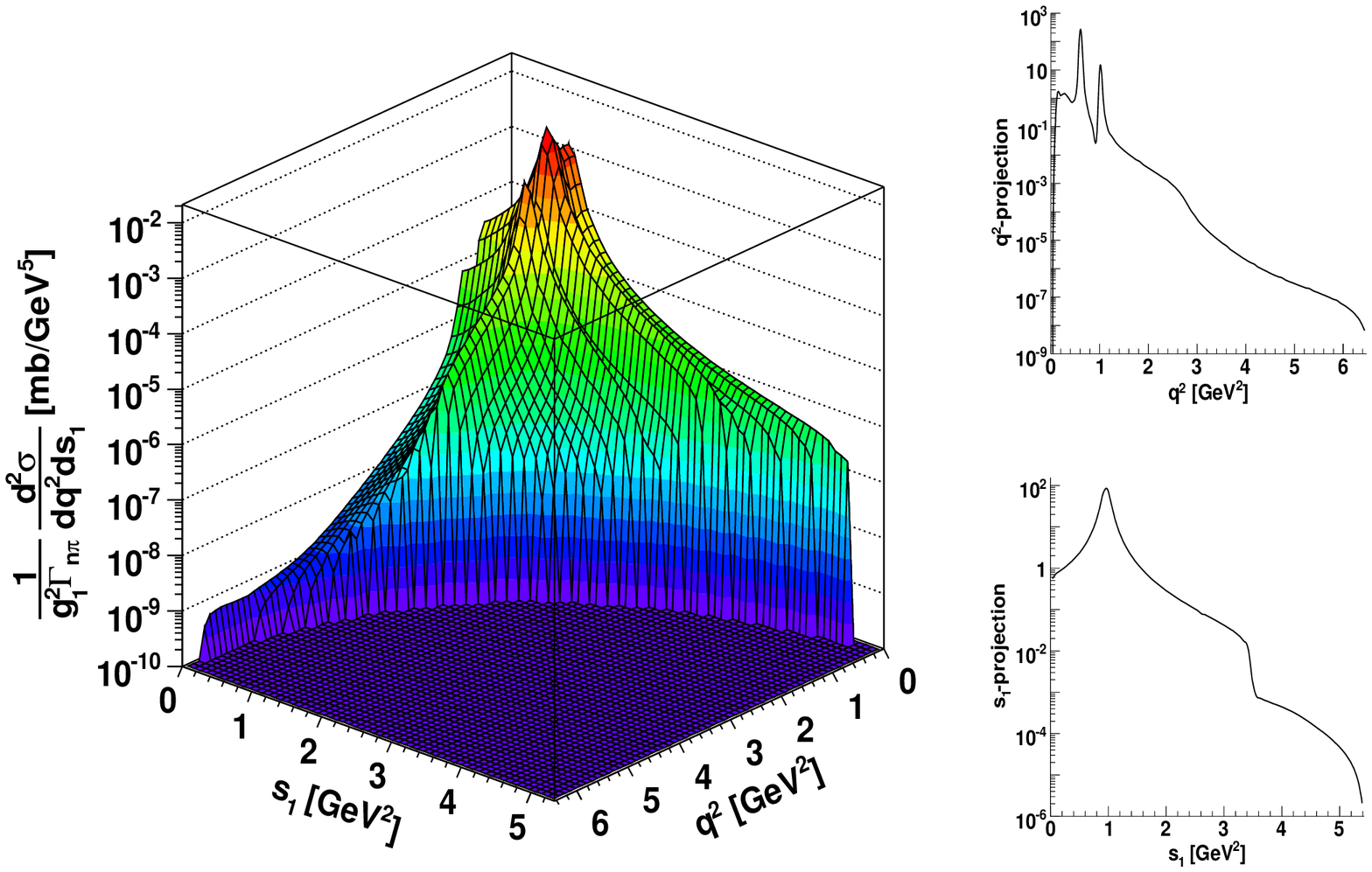}
\caption{\label{Fig:res_sc} (Color online) Double differential cross section for the reaction $\bar p+p\to M^*+n_{\pi}\pi$,  $M^*\equiv f_0(980)$, as a function of $q^2$ and $s_1$.}
\end{center}
\end{figure}

\begin{figure}
\begin{center}
\includegraphics[width=12cm]{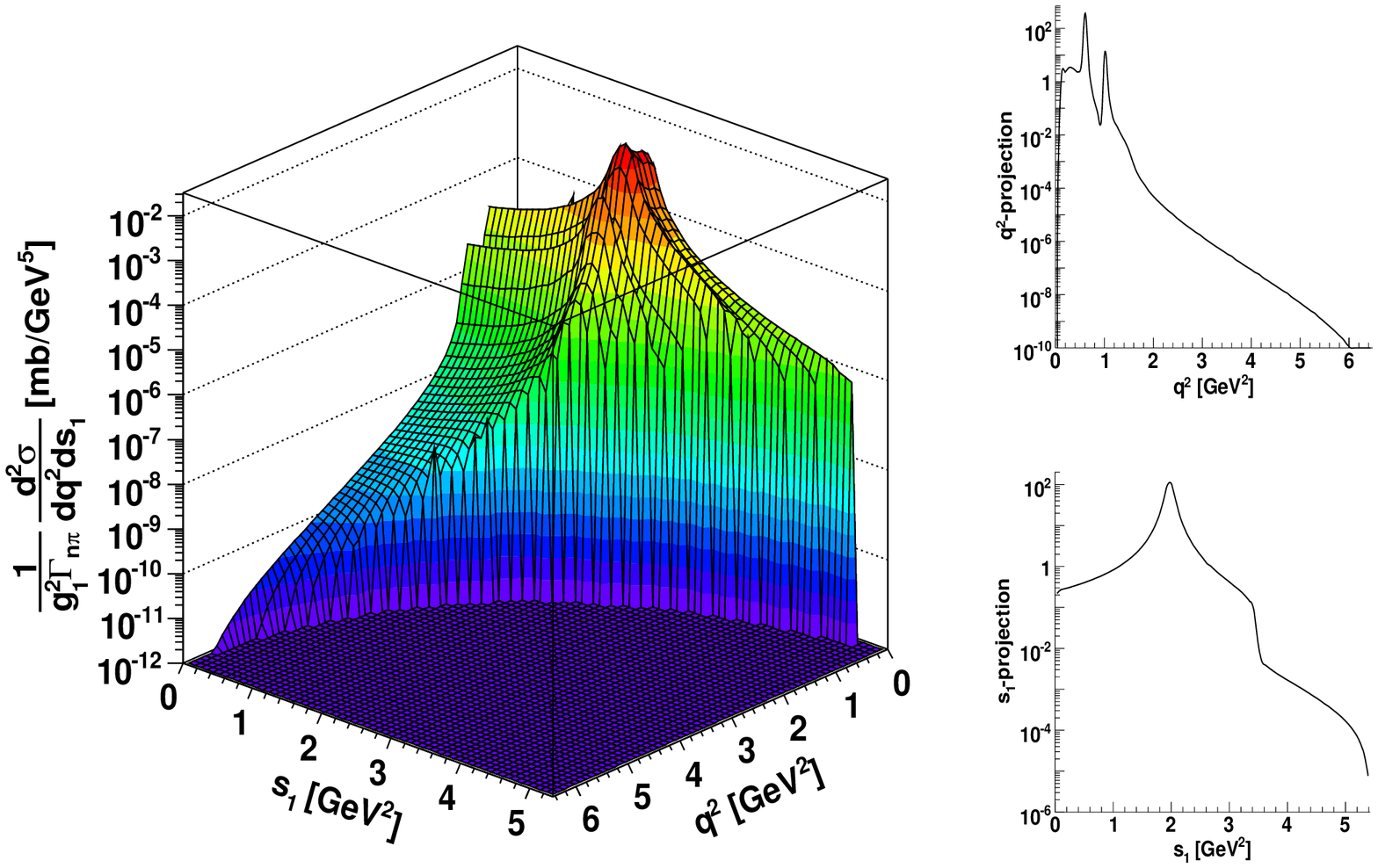}
\caption{\label{Fig:res_psc} (Color online) Double differential cross section for the reaction $\bar p+p\to M^*+n_{\pi}\pi$,  $M^*\equiv \eta(1405)$ as a function of $q^2$ and $s_1$.}
\end{center}
\end{figure}

The differential cross sections for heavy scalar and pseudoscalar states have been calculated for the reaction $\bar p+p\to \ell^+ +\ell^- + n_{\pi}\pi$, as a function of the $s$ and of the total momentum carried by the pions. Since the 
coupling constant $g_1$ and the partial width $\Gamma_{\bar p p}$ are not known, the results for the differential cross section are rescaled by 
these quantities. For the numerical application, we focus on the scalar resonance $f_0(980)$  (Figs \ref{Fig:res_sc}) and on the pseudoscalar resonance, with $\eta(1405)$  (Fig. \ref{Fig:res_psc}). The electromagnetic nucleon form factors have been parametrized according to \cite{Ia73}. 

The two dimensional plots show a structured landscape, limited by the kinematical cut due to the physical accessible region. Peaks appear in the projection to the $q^2$ axis (integration over $s_1$), and are due to the vector meson resonances coming from enhanced photon vector meson coupling included in the FFs model. The projection to the $s_1$ axis  (integration over $q^2$) shows a bump which is due to the excitation of the $p\bar p$ resonant state. 

The absolute value of the cross section is in general about two orders of magnitude smaller than in the case of one pion production, as calculated in Ref. \cite{Ad06}, but still measurable in the resonance region. 

To avoid double counting, we do not consider direct emission of a virtual photon from the resonance, as well as from charged component from its decay, as we imply that it is implicitly taken into account in the resonance width.

Let us discuss now the possibility for the creation of an atomic state of $\bar p N$ type. In the case of a charged atom ($N=n$) the situation is similar to the deuteron case, with a weakly bound state. One may expect several narrow states with masses $\le 2M$ and binding energy of $\simeq 2-3$ MeV. The experimental observation of such states has been reported \cite{Tati}.

The coupling constant $g_1$ related to the creation of the bound state  $(\bar p n)$ can be related (following \cite{AkPo}) to the wave function of this state:
\be
g_1=M^{-3/2}|\Psi(0) |,~ \int |\psi|^2 d^3 r=1,~ \Psi(0) =-M\int _0^{r_0} V(r) r \Psi(r) dr,
\ee
where $\Psi(r)$ and $r_0$ are the wave function and the radius of the  $\bar p n$ bound state and $V(r)$  is the potential of the $\bar p - n$ interaction.

The reaction $\bar{p}+ n\to \pi_1+...+\pi_{n_\pi}$ can be calculated using a  similar formalism with evident modifications, replacing the proton FFs with the  neutron FFs and adding a contact term proportional to $(F_1^n-F_1^p)q_\mu/q^2$. One may expect that the general behavior of the distributions will be similar to the case of $\bar{p}p$ collisions. The size of the relevant resonance should be be much smaller than for $\bar{p}p$ one (several fermi), as for a deuteron-like object.

Neutral atoms, ($N=p$) are of electromagnetic nature, their typical size can be estimated to 25 fm, and their binding energy of the order of 20 keV. The width of these resonances in $S$ state is rather large. Such resonances, with orbital momentum $L=0$ have typically strong interaction width of order of 100 MeV. For $L\ne 0$, the width should be smaller, see Eq. (\ref{eq:eq1}).

In conclusions, we have calculated the double differential cross section for the multipion production in proton antiproton collision, with emission of a leptonic pair. Numerical estimations show that the cross section is measurable in the kinematical region which will be accessible at FAIR and will allow to investigate the formation and the resonant structure of a possible $N\bar N$ system.


\begin{thebibliography}{99}

 
\bibitem{Ad06} C. Adamu\v s\v c\'in, E. A. Kuraev, E. Tomasi-Gustafsson and F. Maas, arXiv:hep-ph/0610429, to appear in Phys. Rev. C.


\bibitem{Bo74}
  L.~N.~Bogdanova, O.~D.~Dalkarov and I.~S.~Shapiro,
  Annals Phys.\  {\bf 84} (1974) 261.
\bibitem{An01}
  A.~V.~Anisovich {\it et al.},
  Phys.\ Lett.\ B {\bf 517} (2001) 273;
  Phys.\ Lett.\ B {\bf 517} (2001) 261.

\bibitem{Tati}
  B.~Tatischeff {\it et al.},
  Phys.\ Rev.\ C {\bf 59} (1999) 1878;
  B.~Tatischeff and E.~Tomasi-Gustafsson,
  arXiv:nucl-ex/0411044. 
 
\bibitem{FAIR} An International Accelerator Facility for Beams of Ions and Antiprotons, {\it Conceptual Design Report}, http://www.gsi.de.

\bibitem{Ba68} 
  V.Baier and V.~S.~Fadin,
  Phys.\ Lett.\  B {\bf 27} (1968) 223;
  G.~Bonneau and F.~Martin,
  Nucl.\ Phys.\  B {\bf 27} (1971) 381;
F. Berends and R. Gastmans,
"Electromagnetic Interactions of hadrons, II, Plenum Press, New York-London, 1978.


\bibitem{AkPo} A.I Akhiezer and I. Pomeranchuk, 'Some questions in hadron theory', Chapter I, Moscow 1950 (in russian).

\bibitem{Ia73}
F.~Iachello, A.~D.~Jackson and A.~Lande,
Phys.\ Lett.\ B {\bf 43}, 191 (1973);  
F.~Iachello,
eConf {\bf C0309101}, FRWP003 (2003) 
[arXiv:nucl-th/0312074];
F.~Iachello and Q.~Wan,
Phys.\ Rev.\ C {\bf 69}, 055204 (2004).

 

  
 
\end{thebibliography}
\end{document}